\documentclass[prb,preprint,showpacs,preprintnumbers,superscriptaddress,amsmath,amssymb]{revtex4}
\usepackage{dcolumn}%
\usepackage{epsfig}
\usepackage{times}
\usepackage{color}
\usepackage{graphicx}

\definecolor{Blue}{rgb}{0,0,1}
\definecolor{NavyBlue}{rgb}{0.14,0.14,0.56}
\definecolor{rot}{cmyk}{0,1,1,0}

\begin{document}
\def\bra#1{\mathinner{\langle{#1}|}}
\def\ket#1{\mathinner{|{#1}\rangle}}
\def\braket#1{\mathinner{\langle{#1}\rangle}}
\def\Bra#1{\left<#1\right|}
\def\Ket#1{\left|#1\right>}
\def\bravert{\egroup\,\vrule\,\bgroup}
\newcommand{\e}[1]{\cdot 10^{#1}}
\newcommand{\wn}{\,cm$^{-1}$}
\newcommand{\ea}{\emph{et al.}}
\newcommand{\dg}{$^{\circ}$}

\title{First-principles calculations of the vibrational properties of bulk CdSe and CdSe nanowires}
\author{M. Mohr}
 \email{marcel@physik.tu-berlin.de}
\affiliation{Institut f\"ur Festk\"orperphysik, Technische Universit\"at Berlin,
Hardenbergstr. 36, 10623 Berlin, Germany}

\author{C. Thomsen}%
\affiliation{Institut f\"ur Festk\"orperphysik, Technische Universit\"at Berlin,
Hardenbergstr. 36, 10623 Berlin, Germany}

\date{\today}

\begin{abstract}

We present first-principles calculations on bulk CdSe and CdSe nanowires with diameters of up to 22\,\AA. 
Density functional linear combination of atomic orbitals and plane wave calculations of the electronic and structural properties are presented and
discussed. We use an iterative, symmetry-based method to relax the structures into the ground state. 
We find that the band gap depends on surface termination.
Vibrational properties in the whole Brillouin zone of bulk CdSe and the zone-center vibrations of nanowires are
calculated and analyzed. 
We find strongly size-dependent and nearly constant modes, depending on the displacement directions.
A comparison with available experimental Raman data is be given.

\end{abstract}

\pacs{ 61.72.uj,
63.22.-m,  
63.22.Gh }

\maketitle

\section{Introduction}

Nanocrystals and nanowires (NWs) have become an important field in solid state physics, especially due to the recent advances in the growth methods.
For example, the fabrication of CdSe NWs or nanorods with small diameter deviations and diameters down to a few
nanometers has been accomplished.\cite{yu03,li01cdse}
On a nanometer scale confinement effects strongly influence the electronic structure. 
While there has been considerable effort in understanding the electronic and optical properties from
theory \cite{yu03,sadowski07} and experiments\cite{chen01,yu03} there have been little efforts to understand the
vibrational properties of such nanostructures and of bulk CdSe, either. 
The phonon dispersion of wurtzite CdSe is only partly available from experiments\cite{widulle99}. 
To our
knowledge there is no first-principles calculation of the phonon dispersion available for hexagonal CdSe. 
Only a bond-charge model calculation with nearest neighbor interactions is available.\cite{Camacho00} Recent
studies on hexagonal materials, like, e.g.  graphite, revealed that for a proper description of the vibrational properties of hexagonal systems interactions up to four
nearest neighbors are needed.\cite{wirtz04}
Confinement effects have been discussed for nanospheres.\cite{trallero98}. Recently, Raman data on small
axial-symmetric CdSe nanorods have been presented.\cite{lange07} 

In the present work we concentrate on the vibrational properties of CdSe bulk and NWs. 
We compare nanowires with diameters between 5 and 22\,\AA. Structural and electronic properties depending on the diameter and surface
termination are investigated. 
We show how the counterparts of the optical bulk modes evolve in frequency with decreasing diameter. We find phonon
modes with higher frequencies than in the bulk, that vibrate mainly on the surface.

\section{Calculational details}
\label{sec:bulk}

\label{sec:calcdetails}

Two different DFT codes were used for the calculation of bulk properties and
NW properties: 
  for the phonon dispersion of bulk material we used linear response, which is
implemented in {\sc ABINIT}.\cite{gonze02,gonze05,abinit,cdseABINIT} %
For large unit cells, as required by one-dimensional systems, however, plane wave basis sets are
computationally very expensive.
Therefore  we used for NWs the {\sc SIESTA} code that uses atom-centered confined
numerical basis functions.\cite{ordejon96,soler02}

\begin{figure}[t]
 \centering
  \epsfig{file=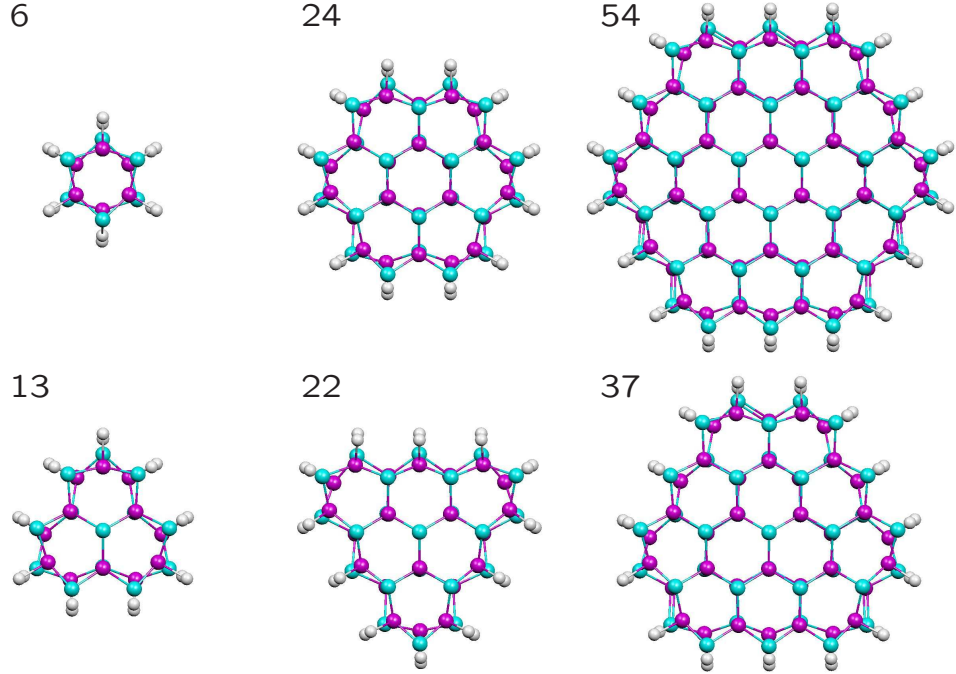,width=0.9\columnwidth}
 \caption{\label{bild:mHhor} (Color online) 
 Unit cells of all calculated hydrogen-coated NWs. The number of
  CdSe-pairs per unit cell is indicated. As can be seen, near the surface, the Cd atoms (magenta, dark) move
 more inwards, as compared to bulk region.\cite{aruguete07}
 }
 \end{figure}
The NWs were generated by cutting appropriate discs out of a wurtzite crystal. The cross section of the relaxed unit cells is shown in
Fig.~\ref{bild:mHhor}. 
We used the local density approximation\cite{perdew81}. Using gradient approximation should not influence the tendencies
will impose. Pseudopotentials
were generated with the Troullier-Martins scheme\cite{troullier91} for the following valence-state configurations: Se
$4s^2(1.89)\quad 4p^4(1.89)$ Cd $5s^2(2.18)\quad 4d^{10}(2.5)$, where the value in parenthesis indicates the
pseudopotential core radii
in bohr. The valence electrons were described by a double-$\zeta$ basis set 
plus an additional polarizing orbital. The localization of the basis followed the standard split scheme and was controlled by
an internal {\sc SIESTA} parameter, the energy shift, for which a value of 50\,meV was used. This resulted in basis functions
with a maximal extension of 3.45\,\AA (Se), 4.13\,\AA (Cd) and 3.2\,\AA (H).
  Periodic images of the NWs were separated by at least 20\,\AA. 
Real space integrations were performed on a grid with a fineness of 0.2\,\AA, which can represent plane waves up to an
energy of 80\,Ry.
For bulk CdSe  a ($6\times 6 \times 4$) Monkhorst-Pack \cite{monkhorst76} mesh in
reciprocal space was used, whereas for wires a minimum of 16 $k$-points  equally spaced along the 1D Brillouin zone was used. 
The phonon calculations were performed with the method of finite-differences.\cite{Yin82}
For the calculation of the LO splitting we used the experimental dielectric high-frequency constant~\cite{verleur67}
$\epsilon_{\infty}^{||}=6.0$ and a modified
version of {\sc SIESTA} described in Ref.\onlinecite{fernandez04}.

The lattice parameters for wurtzite CdSe obtained with {\sc ABINIT} were $a$=4.29\,\AA \ and $c$=7.00\,\AA ; {\sc
  SIESTA} yields $a$=4.34\,\AA \ and
$c$=7.09\,\AA, both in good agreement with the experimental values $a$=4.2999\,\AA \ and $c$=7.0109\,\AA \ from inelastic
  neutron scattering (INS).\cite{widulle99}
We also calculated the bulk modulus and obtained the values 55.58\,GPa ({\sc ABINIT}) and 55.54\,GPa ({\sc SIESTA})
  again in
  good agreement with the experimental value 53.4\,GPa. \cite{bonello93}.

Symmetry considerations helped to reduce the calculation time.
NWs can be described in terms of line groups,\cite{vujicic77}
which are the one-dimensional analogon to space groups. 
The NWs from Fig.~\ref{bild:mHhor} with a hexagon in the
center (upper row) belong to the line group $L(6)_3mc$ which has the 
generator $\{C_6|\mathbf{t}/2\}$. This manifests itself in a screw axis. 
The NWs from Fig.~\ref{bild:mHhor} with three 120\dg \ bonds in the
center (lower row) belong to the line group $L3m$ which has the generator $\{C_3|0\}$. 
In large systems internal stresses can occur, therefore we symmetry based, iterative the relaxation process:
After the NW was relaxed, we generated the whole wire
by applying symmetry operations to a minimal set of atoms in the unit cell. This symmetry generated cell was relaxed again.
 This iterative process was repeated until all forces were below
 0.04 eV/\AA \ after generating the wire. 
Although the relaxation process is considered to not change the symmetry we observed better results with the generating method:
Frequencies of phonons belonging to two-dimensional representations
(degenerate frequencies) were separated by less than 0.05\wn .

\section{Results and discussion}

\begin{figure}[t]
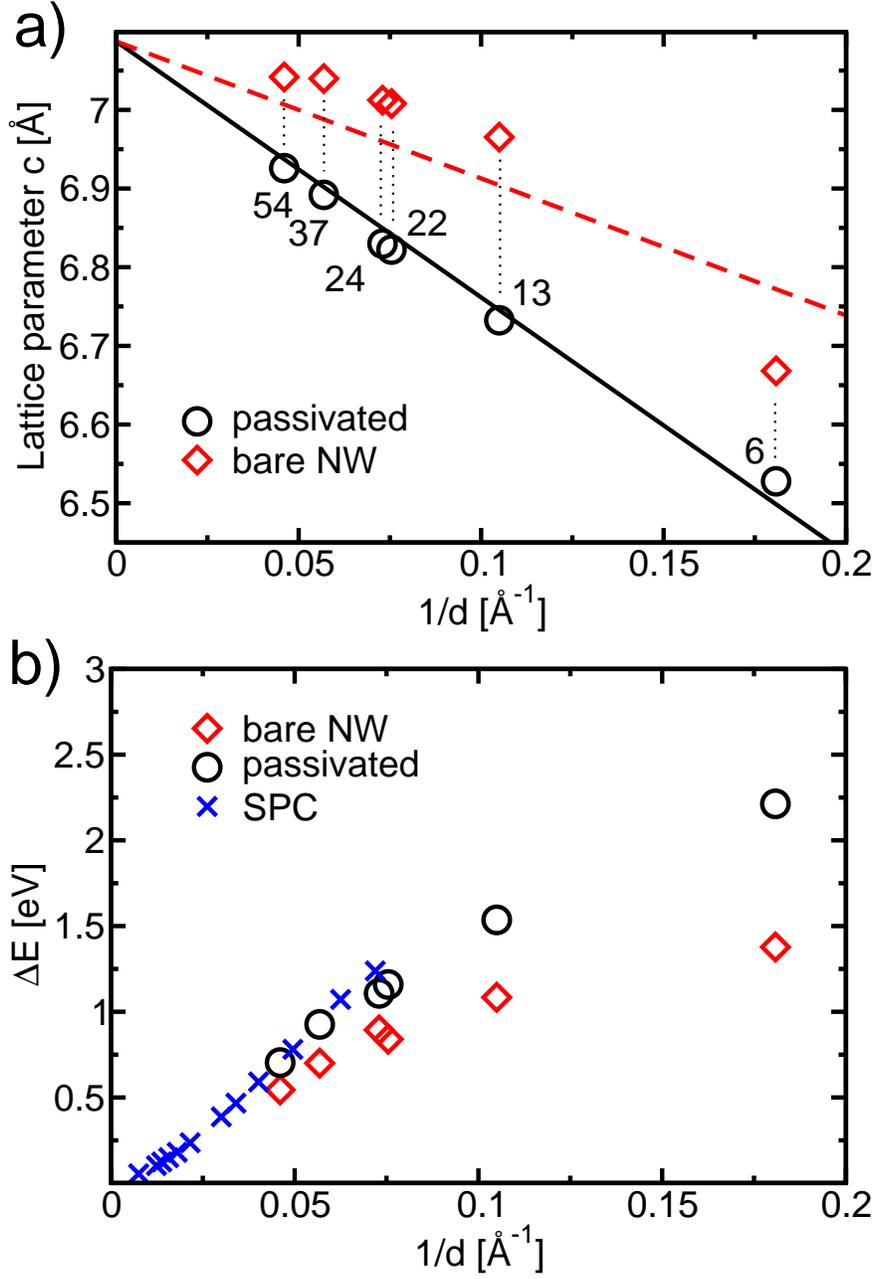

\centering
 \epsfig{file=1d_unitC.eps,width=0.7\columnwidth}
 \epsfig{file=1d_Dgaps.eps,width=0.7\columnwidth}
\caption{\label{bild:1d_Dgaps} 
(Color online) 
a) Evolution of the unit cell lengths of the NWs with and without H-atoms. Bulk CdSe is
  considered as an inverse diameter of 0. The number of CdSe-pairs is indicated.
b) Band gap difference $\Delta E=E_{\mbox{\tiny NW}}-E_{\mbox{\tiny bulk}}$ of the NWs with and
without H-passivation. 
Crosses denote data from semiempirical pseudopotential
calculation (SPC) from Ref.~\onlinecite{yu03}.
}
\end{figure}

The unit cell length $c$ of the NWs changes with diameter and with the addition of a hydrogen passivation shell around them. 
We plotted the unit cell length $c$ %
over the inverse diameter $1/d$ in
Fig.~\ref{bild:1d_Dgaps}. 
The diameter is defined by the average distance of the outermost Se atoms to the rotational axis.
The bare wires have a larger unit cell parameter $c$ than the passivated NWs. The largest increase of 3.5\,\%
was found for the (CdSe)$_{13}$ wire. For the largest wire, the (CdSe)$_{54}$, the increase is
1.7\,\%. We find a linear relationship, which is best described by the
parameters $c_{\mbox{\tiny NW}}(d)=c-m/d$ with the bulk lattice constant $c=7.09$\,\AA \ and $m=3.25$\,\AA$^2$
(passivated) and $m=1.74$\,\AA$^2$ (bare) NWs.

The most dramatic changes in the atomic structure happen near the surface. The surface of all calculated NWs undergo a
reconstruction. This reconstruction was first reported by Ref.~\onlinecite{aruguete07} \emph{via} extended
x-ray absorption fine structure spectroscopy and first-principles calculations.
During this reconstruction the Cd-atoms rotate into the surface to lower the
energy as can be seen in Fig.~\ref{bild:mHhor}. 
The passivation with hydrogen atoms does not qualitatively change this reconstruction.
After relaxation, the Cd-Se bonds show a broad distribution in lengths. In the core the bond lengths deviate from the
bulk value by $<1\,$\% , while those on the surface 
change by up to 5\,\% . The change in bond length causes changes in the force constants which will be discussed later. 

CdSe has a direct electronic band gap. %
It is known from absorption or photoluminescence measurements that the band
gap of CdSe nanocrystallites and NWs shows a strong size dependence.\cite{yu03}
In Fig.~\ref{bild:1d_Dgaps} (lower) we plot the band gap increase (compared to the calculated bulk value) over the inverse
diameter. As can be seen the passivation influences the electronic band structure; it generally increases the band gap
at the $\Gamma$ point. 
At first sight this is a strange behavior as the H-atoms add electronic states to the unit cell. However, at the same time
the lattice parameter $c$ decreases with the addition of the H-atoms which is a stronger effect. Thus the lateral
confinement increases and leads to a net increase of the band gap.
The data from the wires with passivation is in excellent agreement with the shown semiempirical pseudopotential
calculation (SPC) data from Ref.~\onlinecite{yu03} that are based on experimental data.

 \begin{figure}[t]
 \centering
  \epsfig{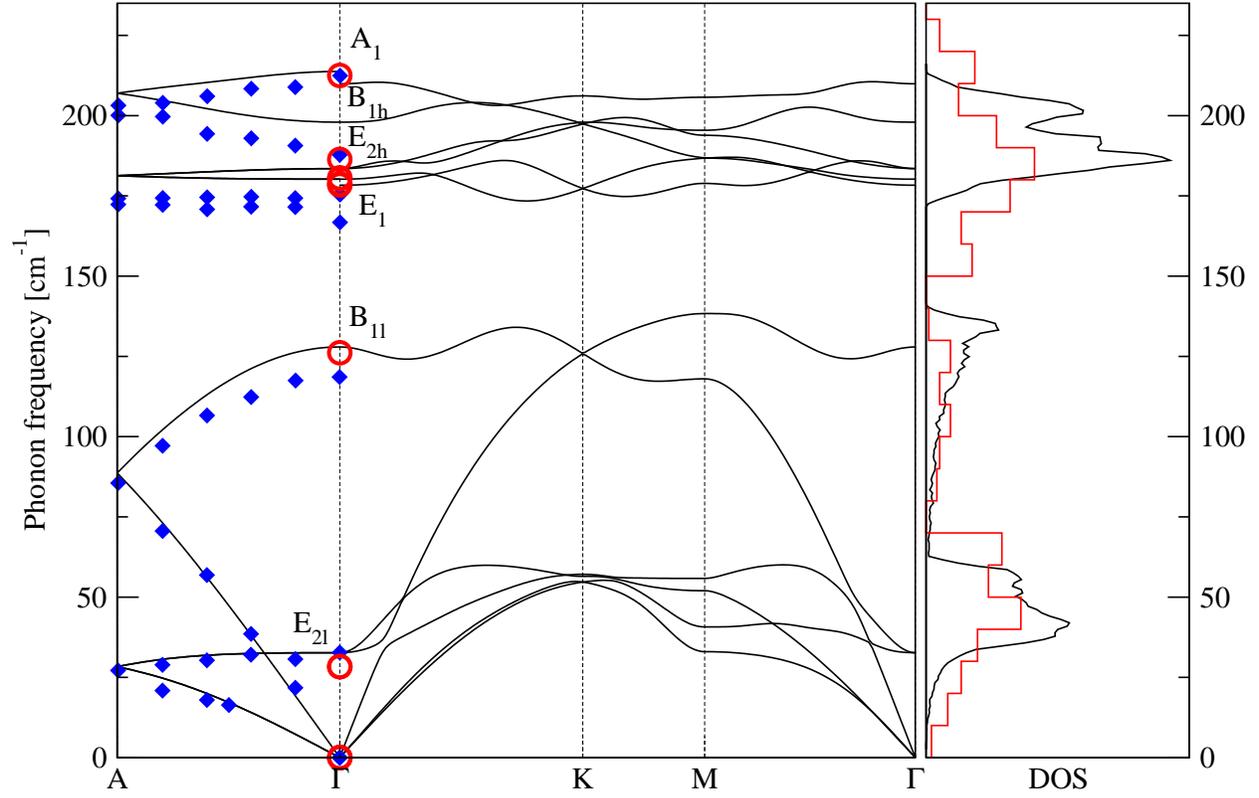}
 \caption{\label{bild:bulkdisp} (Color online) 
Phonon dispersion curves and DOS for bulk CdSe as calculated with ABINIT (solid lines). Diamonds are experimental data
 from Ref.~\onlinecite{widulle99}. Circles are $\Gamma$-point frequencies calculated with SIESTA. In the bulk phonon DOS
(line) we plotted the $\Gamma$-DOS of the ideal (CdSe)$_{54}$-wire (histogram).
 }
 \end{figure}

The phonon dispersion calculated with {\sc ABINIT} is shown in
Fig.~\ref{bild:bulkdisp}. Also shown are experimental frequencies
 obtained with INS along the $\Gamma-$A direction which was multiplied by $\sqrt{(116/112.4)}$ to account for the use of
 $^{116}$Cd isotope instead of natural Cd with an atomic weight of 112.4\,U.\cite{widulle99} 
The $\Gamma$-point phonons calculated with {\sc SIESTA} are also plotted; we compare them later with the NW
frequencies. All frequencies were multiplied by a constant factor to match the experimental bulk LO
frequency at the $\Gamma$-point. 
The dispersion along $\Gamma-K$ and $\Gamma-M$ shows an overbending near the $\Gamma$-point, which can be found in other
wurtzite crystals as well.\cite{ruf01,serrano04} 
When treating the vibrations in the zone-folding approximations, it is important to know whether or not there is
overbending.\cite{richter81}

We have plotted a normalized histogram of the $\Gamma$-phonons of a representative NW
into the bulk phonon density of states (DOS) in Fig.~\ref{bild:bulkdisp}. Acoustic modes below
$75$\wn are shifted to higher frequencies, whereas the optical modes around 200\wn are red-shifted, 
 a general trend, \emph{e.g.}, as reported by Sun in Ref.\onlinecite{sun07}.
Additionally, optical modes in NWs are found with frequencies higher than the LO-mode in the bulk: These modes vibrate mainly on the
surface, their frequency exceeding the bulk LO by at most 15\wn . The reason for the increase lies in
the higher coordination of the reconstructed Cd atoms as an inspection of the force constants for the outermost Cd atoms
for the slab shows. The sum of the forces on this atom when slightly displaced is higher by 4\,\%  than that of an Cd
  atom in in the center for equal displacement.
Force constants of atoms lying in the 2nd layer or closer to the rotational axis do not significantly differ from those in the bulk.
This is similar to the subsurface modes predicted for the zirconium surface.\cite{yamamoto96}

\begin{figure}[t]
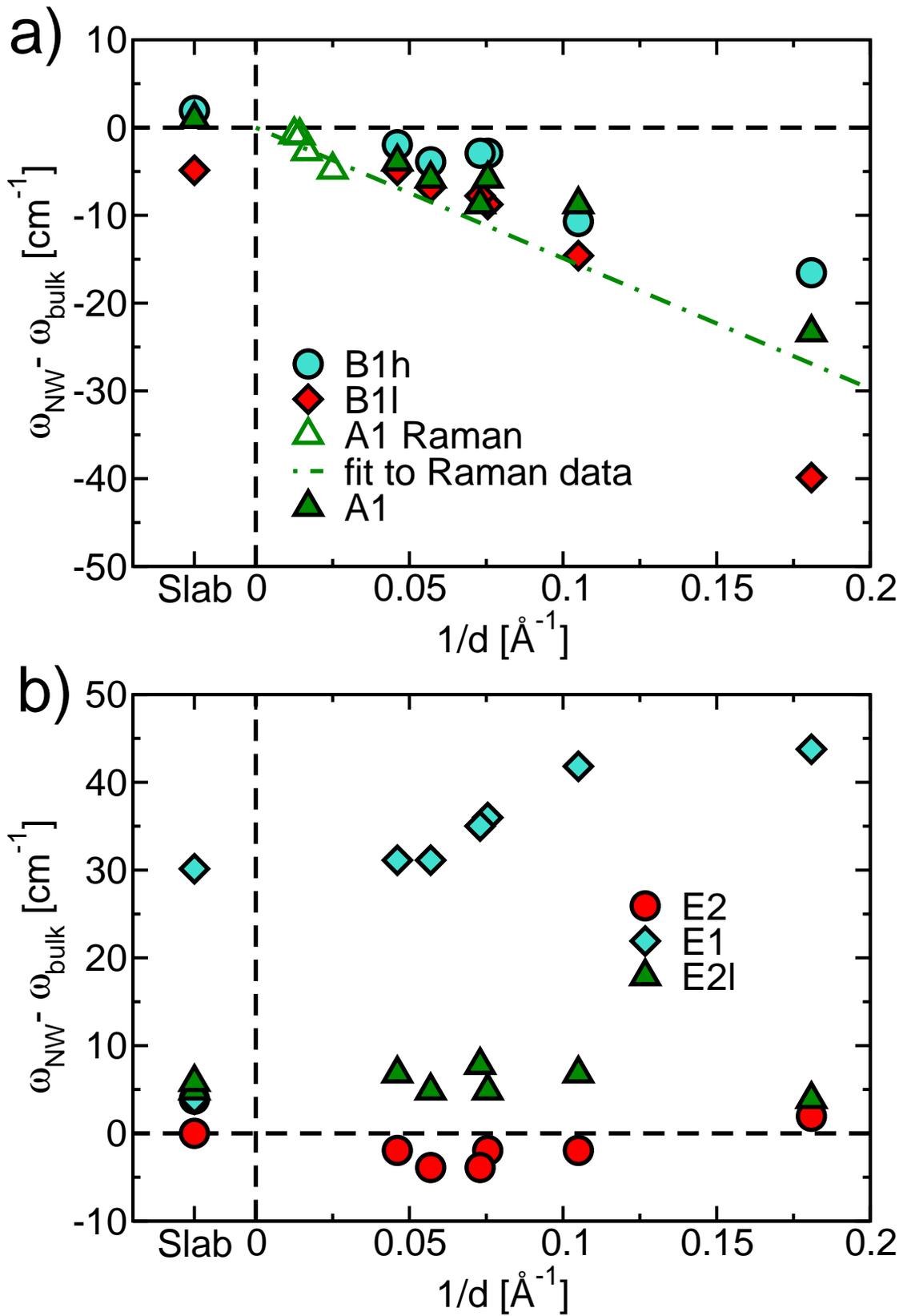

 \centering
  \epsfig{file=1d_1Dim.eps,width=0.9\columnwidth}
  \epsfig{file=1d_2Dim.eps,width=0.9\columnwidth}
 \caption{\label{bild:1d_allemoden} (Color online) 
Phonon energy difference $\omega_{\mbox{\tiny NW}}-\omega_{\mbox{\tiny bulk}}$ over inverse diameter for all optical bulk modes. On the left are
 the modes of a slab with a thickness of 20\,\AA \ which corresponds to the diameter of the biggest calculated NW. a)
 Phonons belonging to 1-dimensional representations. b) Phonons belonging to 2-dimensional representations. 
 }
 \end{figure}

We now focus on optical phonon modes that exist in bulk material and their counterparts in NWs. 
The evolution in phonon frequency with changing NW diameter is shown in Fig.~\ref{bild:1d_allemoden}, where we plot
the phonon-frequency difference between
the NW and bulk over the inverse diameter.
All considered bulk modes that belong to one-dimensional representations show a strong systematic redshift for reduced diameters  (upper
plot).\cite{sun07} In contrast,
the modes belonging to two-dimensional representations  show a smaller size dependence (lower plot). The rigid shift of the $E_1$-mode will be explained later.
This different size dependence has also been predicted for Si NWs in Ref.~\onlinecite{yangphd}. 
They find the LO mode with eigenvectors parallel to the NW axis shows a stronger size dependence, than the two TO modes
with eigenvectors perpendicular to the axis.
This is the same here, all modes belonging to one-dimensional representations have eigenvectors almost parallel to the
wire axis and depend strongly on size.

Polar semiconductors are known to have different frequencies for LO and TO modes at the $\Gamma$-point. The reason for
this is that long-wavelength optical longitudinal atomic vibration induce a polarization field that contributes to the
restoring forces due to the long-range Coulomb interaction. 
It has been shown, that this Coulomb interaction can be treated separately from the short ranged dynamical calculation. 
In detail the 
dynamical matrix can be split into an analytical and a non-analytical part, the latter containing the long-range Coulomb
forces. Its evaluation is performed as an Ewald summation.\cite{born_huang, cochran62, giannozzi91}
This non-analytical part depends on the unit cell volume and the dielectric high-frequency constant $\epsilon_\infty$.
In bulk CdSe with an experimental dielectric constant\cite{verleur67} of $\epsilon_{\infty}^{||}=6.0$ a
splitting of 32\,cm$^{-1}$ is obtained in good agreement with the experimental value of 39\wn.\cite{widulle99}
We apply the same model to NWs and take the wire cross section times unit cell height as cell volume. To account
for the one-dimensionality we only sum over k-vectors along the wire axis. The same dielectric constant
$\epsilon_{\infty}^{||}$ is used.\cite{mahan06}
We obtain a splitting of only a few \wn \ for larger wires; much smaller than in the bulk.
We attribute the underestimated splitting to the fact that the displacements within the LO mode in the NW are not well
ordered. They are parallel only in the NW-core and even 
antiparallel near the surface. Consequently
the polarization as sum of dipoles becomes weaker compared to bulk. 
In Fig.~\ref{bild:1d_allemoden} the energy difference of the $A_1$-modes is calculated without any contribution from the
long-ranged Coulomb interactions, neither for bulk nor NWs.
The bulk frequency reference becomes then 177\wn
 (calc. with {\sc SIESTA}).  %
 For this mode experimental Raman data is available for nanorods with aspect ratios
of $>5$ and  diameters between 4 and 8\, nm.\cite{lange07,lange08}
A linear fit $\omega_{\mbox{\tiny NW}}-\omega_{\mbox{\tiny bulk}}=m/d$ leads to $m=-149$\wn\AA. 
The calculated frequencies of the $A_1$-modes shown in Fig.~\ref{bild:1d_allemoden} are in good agreement with the
extrapolated experimental fit.

A rigid shift of the $E_1$-mode was predicted by 
Fuchs and Kliewer \cite{fuchs65} in a calculation of an ionic crystal slab. They showed that phonons with a
polarization perpendicular to the slab surface have frequencies similar to the LO-mode in the bulk.
This is attributed to surface charges. A characteristics of this  mode is that the Cd and Se sub-lattices move into opposite directions. 
To verify this we calculated the phonon frequencies of a thin slab extending into the
$y$ and $z$-directions. The non-analyticity is taken along $z$.  We find that the
degeneracy of the E$_1$-mode is lifted into two modes: One has a polarization perpendicular to the surface and a frequency of
the LO in the bulk; the other one has a polarization vector parallel to the surface and a frequency of the E$_1$-mode in
bulk.
The rigid shift of the $E_1$ mode in NWs is thus caused by surface charges in a one-dimensional system and the
mode is
degenerated again, but now has the frequency of the LO mode in the bulk. 
The $E_1$-modes in Fig.~\ref{bild:1d_allemoden}~b) show a size dependence. This can be qualitatively explained considering the
electric field generated by the surface charges, that becomes weaker for larger diameters.

\section{Summary}\label{sec:summary}
We presented first-principles calculations of CdSe bulk and NWs. 
We used an iterative relaxation process that reproduces well the degenerate phonon modes.
The effect of a H-passivation on NWs resulted in smaller unit
cells and larger band gaps.
We analyzed the vibrational properties of NWs. A reconstruction in the surface leads to higher force constants, resulting
in surface modes with frequencies up to 15\wn \ higher than in the bulk. 
The size dependence of phonon modes that correspond to bulk optical $\Gamma$ modes was analyzed. 
We find that modes which vibrate mainly along the wire axis showed a stronger size-dependence than
modes with eigenvectors perpendicular to the axis.

\section{Acknowledgements}
The authors would like to thank M. Mach\'{o}n, H. Scheel and S. Reich for useful discussions;
J. M. Pruneda for supplementing the program to calculate the nonanalytic part; H. Lange for providing experimental
data prior to publication.


\begin{thebibliography}{38}
\bibitem[{cds()}]{cdseABINIT}
\bibinfo{note}{We used the LDA functional of Perdew and Wang in {\sc
  ABINIT}.\cite{perdew92} The valence electrons were expanded in a plane wave
  basis with an energy cutoff of 30 Hartree. The $4d$ electrons of Cd were
  taken as valence. Core electrons were replaced by Troullier-Martins
  pseudopotentials.\cite{troullier91} A shifted $10\times10\times4$ sampling
  grid was used for the integration over the Brillouin zone. The dynamical
  matrix was calculated on a $10\times10\times4$ $q$-grid. Force constants were
  obtained via a Fourier transformation and interpolated to obtain phonons at
  arbitrary points in the Brillouin zone. The splitting of the LO and TO
  phonons was calculated with the standard model.\cite{cochran62, giannozzi91}
  {\sc ABINIT} yields a dielectric high-frequency constant of
  $\epsilon_{\infty}^{||}=7.2$}.
\end{thebibliography}
\end{document}